\documentclass[sigconf]{acmart}

\usepackage{booktabs}
\usepackage{multirow}
\usepackage{xcolor}
\usepackage{graphicx}
\usepackage{subcaption}
\usepackage{balance}
\usepackage{enumitem}
\usepackage[most]{tcolorbox}
\usetikzlibrary{positioning,calc,arrows.meta}

\newcommand{\RQone}{RQ\textsubscript{1}}
\newcommand{\RQtwo}{RQ\textsubscript{2}}
\newcommand{\RQthree}{RQ\textsubscript{3}}
\newcommand{\RQoneTitle}{How much previously correct behavior is lost as requirements accumulate across turns in LLM coding conversations?}
\newcommand{\RQtwoTitle}{What multi-turn-specific bugs emerge, and how are they distributed across taxonomy categories and requirement change types?}
\newcommand{\RQthreeTitle}{To what extent can interaction-level mitigation strategies preserve prior behavior in multi-turn coding conversations?}

\setcopyright{none}
\acmConference[ASE '26]{41st IEEE/ACM International Conference on Automated Software Engineering}{October 2026}{Munich, Germany}

\begin{document}
\raggedbottom

\title{Regression Accumulation in Multi-Turn LLM Programming Conversations}

% \begin{abstract}

\author{Yonghui (Andie) Huang}
\affiliation{%
\
  \institution{ Massey University}
  \country{New Zealand}
}
  \email{andie.huang@massey.ac.nz}

\author{Lin Ma}
\affiliation{%
\
  \institution{Alibaba Research Center for Complexity Sciences, Hangzhou Normal University}
  \country{China}
}
  \email{malin@hznu.edu.cn}

\author{Amjed Tahir}
\affiliation{%
\
  \institution{Massey University}
  % \city{Palmerston North}
  \country{New Zealand}
}
  \email{a.tahir@massey.ac.nz}

\author{Qian Zhang}
\affiliation{%
\
  \institution{University of Otago}
  % \city{Palmerston North}
  \country{New Zealand}
}
  \email{zhaqi075@student.otago.ac.nz}

\author{Liwen Xiao}
\affiliation{%
\
  \institution{University of Otago}
  % \city{Palmerston North}
  \country{New Zealand}
}
  \email{xiali449@student.otago.ac.nz}

\author{Lysa Xiao}
\affiliation{%
\
  \institution{University of Otago}
  % \city{Palmerston North}
  \country{New Zealand}
}
  \email{lysa.xiao@postgrad.otago.ac.nz}

\begin{abstract}
In LLM-assisted software development, coding is often iterative.
In this paper, \emph{single-turn} refers to settings where the LLM model responds in one request with one code suggestion, whereas \emph{multi-turn} refers the model responds across multiple sequential requests within the same conversation. 
The \emph{regression accumulation} refers to the later code suggestion break the requirements introduced in earlier turns.
Reliability depends not only on satisfying the current turn, but also on preserving requirements in earlier turns. 
Yet, there are limited evaluations of the accumulation of regression in multi-turn coding conversations.

To address this gap, we conduct an empirical study of multi-turn regression in LLM programming. 
We construct 542 tasks from HumanEval+ and MBPP+ benchamrks, construct each task with an 8-turn requirement-evolution chain. 
We evaluate six LLM models on 26,016 turn instances ($542 \times 6 \times 8$). 
At each turn, we check if the current code suggestion still passes earlier benchmark tests, and we analyze 384 failure cases drawn from the failure population in our 26,016 turn instances to derive and validate a taxonomy of multi-turn regression bugs through independent four-annotator labeling.
Our results show that regression accumulation appears across all six models, with 40\% to 73\% of tasks losing previously correct behavior over the full conversation. 
Final-turn quality is lower than initial-turn quality across models, and especially in turns adding input validation or broader input types. 
Manual analysis shows that \emph{Cross-Turn Conflict}, where later code conflicts with earlier requirements, is the main failure class.
We further find that \emph{Verification Gate}, which checks new code against prior tests and triggers rollback and retry, is the only strategy that consistently improves all models, raising final-turn quality from 75.8\% to 87.9\% on DeepSeek-V3 and from 31.6\% to 47.3\% on Llama-3.1-8B.

These findings suggest that strong performance on single-turn tasks can overestimate reliability in multi-turn coding conversations. Future evaluation and tool design should test whether later code suggestions preserve requirements from earlier turns and should include \emph{Verification Gate} mechanisms.
\end{abstract}

% \end{abstract}


\maketitle

\section{Introduction}\label{sec:intro}

Software programs are rarely developed in a single step. They evolve through repeated modifications that add features, broaden input handling, restructure interfaces, or improve performance. A central engineering risk in this process is \emph{regression}, in which a change breaks behavior that previously worked correctly. Regression testing is the typical response to this risk~\cite{yoo2012regression}%, and classic software evolution research explains why the problem persists. 
Both early and later software evolution research suggest that continual modification can increase system complexity and maintenance risk over time~\cite{belady1976model,fluri2007change,bohner2002impacts}.

AI code-generation systems further highlight this maintenance issue. Developers increasingly use these systems in interactive coding conversations rather than relying only on one-shot prompts~\cite{hao2024shared}. In a \emph{single-turn} setting, the model answers one prompt with one code suggestion. In a \emph{multi-turn} setting, the model continues the same conversation across several turns and produces a new code suggestion after each follow-up instruction. In such sessions, each turn may address a local requirement, but the current code suggestion must remain consistent with the code suggestions and requirements established in \emph{earlier turns}~\cite{belady1976model,yoo2012regression}.

Current evaluation practice does not capture this risk well because it is misaligned with how these systems are actually used. In practice, developers increasingly rely on multi-turn coding conversations, where each new code suggestion is built on what earlier turns already established. By contrast, single-turn benchmarks such as HumanEval~\cite{chen2021evaluating} and MBPP~\cite{austin2021program} ask whether a model can produce a correct code suggestion for one prompt. Recent multi-turn benchmarks~\cite{rawal2025mtsec, wang2025codeif, codeflowbench2025, laban2025llms} show that models are less likely to produce a code suggestion that satisfies the latest instruction once requirements are distributed across several turns in the same conversation. However, they still focus mainly on whether the code suggestion at the current turn satisfies the latest instruction. They do not directly test whether requirements established in earlier turns are still preserved after later changes. This is not a problem with coding tasks themselves. It is a limitation of what current evaluation practice chooses to measure. As a result, a model can appear successful at turn $t$ while silently breaking requirements introduced at turns $1$ to $t-1$. This mismatch can lead us to overestimate reliability in realistic workflows and leave users to recover broken functionality through extra debugging and repair. We therefore need to understand how much regression accumulates across turns, what kinds of failures dominate it, and which interaction-level controls can reduce it.
%Current evaluation practice does not capture this risk. Single-turn benchmarks such as HumanEval~\cite{chen2021evaluating} and MBPP~\cite{austin2021program} measure whether a model can produce a correct initial solution. Recent multi-turn benchmarks~\cite{rawal2025mtsec, wang2025codeif, codeflowbench2025, laban2025llms} show that performance drops once requirements are distributed across several turns. However, they still focus on whether the \emph{current} turn is satisfied. They do not directly test whether functionality established in \emph{earlier} turns survives after later changes. A model can therefore appear successful at turn $t$ while silently breaking behavior established at turns $1$ to $t-1$.

We still lack an evaluation of whether later-turn code suggestions preserve requirements established in earlier turns. This paper addresses that gap by studying whether AI-generated code suggestions preserve requirements established in earlier turns. We construct 8-turn evolution chains and evaluate each code suggestion against the benchmark tests.
We conduct this study on 542 tasks from HumanEval+ and MBPP+ using GPT-4o, DeepSeek-V3, Qwen2.5-Coder-32B, Qwen3-32B, DS-R1-Distill-32B, and Llama-3.1-8B, for a total of 26,016 turn instances ($542 \times 6 \times 8$). Our results show regression accumulation across all model families, with the steepest losses after error-handling and input-generalization requests. Our manual analysis shows that these failures more often reflect incompatibilities between new code suggestions and earlier-turn requirements than simple context loss. We also find that mitigation strategies such as \emph{Verification Gate}, which checks each new code suggestion against earlier-turn tests and allows one retry from the last passing one, are effective when multi-turn coding conversations explicitly verify whether new code suggestions preserve earlier-turn requirements.

\smallskip
\noindent The main contributions from this work are:
\begin{enumerate}[leftmargin=*,nosep]
\item We present \textbf{a benchmark construction method and executable evaluation protocol} for studying regression accumulation in multi-turn LLM programming. The protocol combines 542 HumanEval+ and MBPP+ tasks with 8-turn evolution chains and a fixed regression oracle derived from benchmark-provided tests.
\item We conduct \textbf{a systematic empirical study} across six LLMs and 26,016 turn instances to analyze how regression accumulation evolves across turns, model families, and requirement change types.
\item We develop \textbf{a multi-turn bug taxonomy} with eight multi-turn bug types in three classes, plus Baseline Failure for Turn~1, from manual analysis of 384 failure instances and four-annotator labeling.
\item We formulate and evaluate \textbf{interaction-level mitigation strategies} for multi-turn coding, including a verification-based rollback-and-retry design for code suggestions that introduce regression.
\end{enumerate}

\section{Related Work}\label{sec:related}

\subsection{Software Evolution and Regression Under Change}

Regression under change has long been a central concern in software engineering. As systems evolve, new functionality must be integrated without breaking behavior that already works. Classic software evolution research explains why this risk persists. Lehman and Belady~\cite{belady1976model} argue that software subject to continual modification tends to accumulate complexity and quality risk. Maintenance research commonly distinguishes corrective, adaptive, perfective, and preventive changes~\cite{lientz1980software}. Other work classifies the concrete edits developers make during software evolution~\cite{fluri2007change}. Regression testing operationalizes this concern by checking whether existing behavior survives new modifications~\cite{yoo2012regression}. We adopt this lens for conversational code generation. In our setting, each turn acts as a maintenance step, and the central question is whether the generated update preserves prior behavior.

\subsection{Interactive and Multi-Turn LLM Coding}

Recent surveys and perspective papers position LLM-based development as a growing software engineering research area~\cite{hou2024llm4se,gao2025challenges}. Empirical work also shows that developers use LLM assistants through shared, multi-turn interactions in collaborative settings such as pull requests and issues~\cite{hao2024shared}. These studies motivate evaluating coding assistants in iterative workflows rather than only through one-shot generation.

Recent benchmarks have begun to study multi-turn code generation directly. MT-Sec~\cite{rawal2025mtsec}, CodeIF-Bench~\cite{wang2025codeif}, CodeFlowBench~\cite{codeflowbench2025}, and CodeAssistBench~\cite{codeassistbench2025} all evaluate code generation across sequences of instructions. Outside software tasks, Laban et al.~\cite{laban2025llms} show that multi-turn performance declines across a broad range of domains. These studies establish that turn history matters, but most code benchmarks still score each turn against tests written only for the current request. That setup captures turn-level instruction following. It does not reveal whether later edits preserve prior behavior.

Our protocol instead keeps the benchmark's original tests as a fixed regression oracle at every turn, making regression directly observable. A model can satisfy the latest instruction while still breaking behavior introduced earlier in the conversation. We build on HumanEval~\cite{chen2021evaluating}, MBPP~\cite{austin2021program}, and their enhanced variants HumanEval+ and MBPP+~\cite{liu2024evalplus}, because their validated tests support controlled longitudinal evaluation under repeated change.

\subsection{Failure Modes and Repair in LLM-Generated Code}

Studies of LLM code suggestions have cataloged common single-turn failure modes such as incorrect logic, missing edge cases, and API misuse~\cite{abbassi2025taxonomy}. A separate line of work investigates iterative repair through self-debugging, execution feedback, and learned self-correction~\cite{chen2024selfdebug,kumar2025score,gehring2025rlef,shi2025mgdebugger,jin2026reveal}. These systems aim to improve one candidate solution through repeated repair attempts.

We study a different failure setting. The model is not revising the same task until it passes. It is responding to a sequence of distinct change requests whose effects must compose over time. This difference exposes bug patterns that do not appear in single-turn evaluation, including omission of earlier-turn requirements and failures caused by accumulated code changes. In our paper, prior work on long-conversation drift~\cite{dongre2025driftnomore} motivates recap-based prompting that restates prior requirements, while one-step recoverability~\cite{jain2025mucode} motivates a rollback-and-retry strategy that recovers from code suggestions that introduce regression using the last known-good code.

\section{Study Design}\label{sec:approach}

\begin{figure*}[t]
\centering
\includegraphics[width=\textwidth]{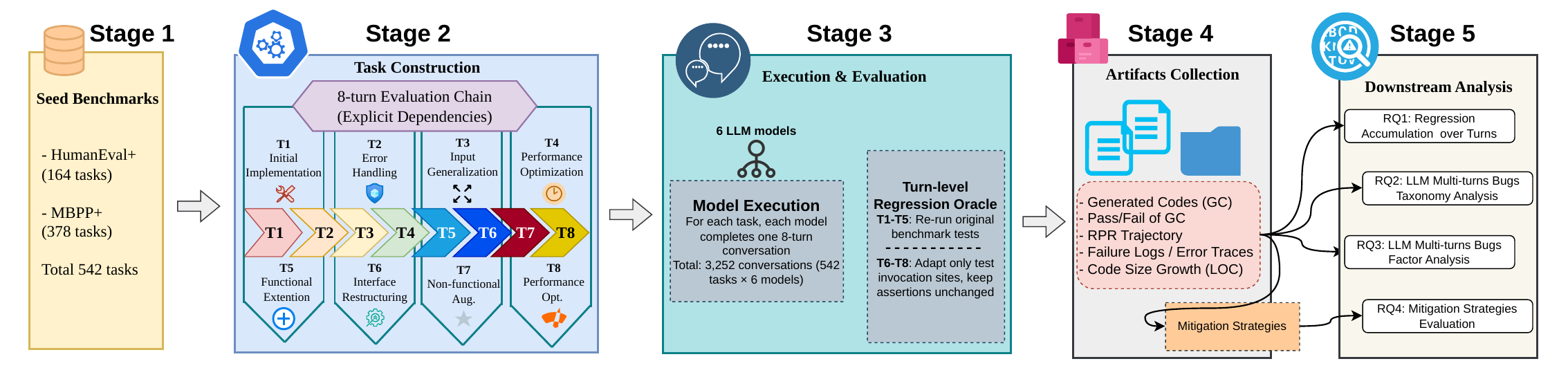}
\caption{Overall study design and data flow. Seed tasks are instantiated as 8-turn evolution chains, executed once for each task on each model, and evaluated with the same benchmark-provided tests at every turn.}
\Description{An overview figure showing the study pipeline from benchmark seed tasks, to 8-turn evolution chain construction, to model execution, to turn-level regression evaluation, to collected artifacts, and finally to the four research-question analyses.}
\label{fig:overview-pipeline}
\end{figure*}

Our study was carried out in five stages. We begin with 542 Python seed tasks drawn from two established benchmarks: 164 Python function-level tasks from HumanEval+ and 378 Python function-level tasks from MBPP+. We include all tasks from both benchmarks and combine them into a single seed-task pool by applying the same fixed 8-turn prompt structure and benchmark-provided tests to every task. We then convert each task into a fixed 8-turn programming conversation. At Turn~1 (T1), the model is asked to produce an initial code suggestion from the benchmark's original task prompt. For Turns~2 to 8 (T2 to T8), we define controlled requirement changes using six software-maintenance change types and instantiate them with fixed prompt templates, each designed to modify code suggestions from earlier turns while preserving previously established requirements. Next, we execute this conversation once for each task on each model. After each turn, we re-run the benchmark tests to measure whether requirements established in earlier turns are preserved. This process yields per-turn code suggestions, test outcomes, turn-by-turn regression results, and the failing code suggestions, failed tests, and error messages returned by Python execution or assertion failures that are used in the downstream analyses for \RQone{} to \RQthree{}. Figure~\ref{fig:overview-pipeline} summarizes this workflow and data flow.

\subsection{Task Construction}\label{sec:construction}

We source 542~Python seed tasks from two established benchmarks, HumanEval+~\cite{liu2024evalplus} (164 Python function-level tasks) and MBPP+~\cite{liu2024evalplus} (378 Python function-level tasks). Both benchmarks are widely used in LLM code generation research and provide rigorously validated test suites. We include all tasks regardless of complexity, using the complexity of the benchmark reference solution as a control variable rather than a selection criterion.

To account for differences in seed-task size, we use the lines of code (LOC) in the benchmark reference solution as a coarse structural proxy~\cite{nagappan2005static, herzig2013predicting}. We do not treat LOC alone as a complete measure of software complexity. Rather, we use it only to stratify these short function-level benchmark tasks by reference-solution size. Based on this measure, we group tasks into three strata: \emph{easy} (1--3 lines, $n$=322), \emph{medium} (4--8 lines, $n$=162), and \emph{hard} ($\geq$9 lines, $n$=58). We also consider a complementary notion of task difficulty for the models themselves. Specifically, we use Turn~1 (T1) pass rate to capture how solvable each task is at the initial code-suggestion turn, so that we can analyze how initial solvability interacts with subsequent regression accumulation.

Each seed task is instantiated as the same fixed 8-turn conversation by filling a small set of benchmark-derived fields into the fixed turn templates introduced later in this section. These fields include the target function name (\texttt{\{entry\_point\}}) and, for the restructuring turn, the derived class name (\texttt{\{class\_name\}}). This deterministic instantiation keeps prompt construction consistent across tasks and avoids introducing additional variation from task-specific prompt writing.

\smallskip
\noindent\textbf{Regression Test Strategy.}
A key challenge arises at Turn~6 (T6) because, unlike earlier turns, it changes how the benchmark tests call the code, \textit{i.e.,}~from standalone functions to class methods. To preserve the same benchmark-provided tests across turns, we continue to use the benchmark's original test suite throughout the conversation. For Turns~1 to 5 (T1 to T5), the tests are executed unchanged. For Turns~6 to 8 (T6 to T8), we deterministically rewrite only the invocation sites by prepending object instantiation and replacing calls of the form \texttt{f(args)} with \texttt{\_obj.f(args)}. The assertions themselves are left unchanged. This design keeps the evaluation tied to the benchmark's validated tests while preventing T6 failures from being counted merely because the required refactoring changes the call interface.

\subsection{Cross-Turn Dependency Design}\label{sec:coupling}

Existing multi-turn benchmarks often decompose tasks into independent subtasks~\cite{rawal2025mtsec} or separately verifiable instructions~\cite{wang2025codeif}. Such designs are useful for measuring whether a model can satisfy the requirement stated at the current turn. However, they are less suitable for studying regression accumulation because a later turn can be solved without modifying or preserving code suggestions from earlier turns. In that setting, a model may succeed at the current turn even if it would fail to maintain earlier-turn requirements under continued code evolution. Our goal is different. We want later turns to modify code suggestions produced earlier in the same conversation and to preserve established requirements. This design choice is consistent with software maintenance settings in which later modifications depend on and interact with earlier ones~\cite{belady1976model,fluri2007change,yau1978ripple,bohner2002impacts}. We therefore construct each subsequent turn so that it must be applied on top of the current code suggestion rather than as an independent new solution. For example, Turn~4 (T4) adds caching to the code suggestion that already includes Turn~2 (T2) validation and Turn~3 (T3) type generalization, and Turn~6 (T6) refactors the accumulated functionality from Turns~1 to 5 (T1 to T5) into a class while preserving the same requirements. Concretely, if T2 requires the code suggestion to reject invalid inputs and T3 requires it to accept string representations of numeric inputs, then a later turn such as T4 must add caching without removing either the validation introduced in T2 or the input conversion introduced in T3.

\subsection{Requirement Change Types}\label{sec:change-types}

Turn~1 is treated separately as the initial code-suggestion baseline. For Turns~2 to~8, we derive a pool of maintenance-like follow-up changes from prior software-maintenance, software-evolution, and source-code-change literature~\cite{lientz1980software,chapin2001types,fluri2007change}. We therefore select six change types for benchmark construction based on three benchmark-design requirements. Each change type should represent a common follow-up modification, create explicit preservation obligations, and be instantiable consistently across many benchmark tasks. We do not claim that these six types exhaust all realistic maintenance turns. Rather, they provide a controlled subset for reproducible comparison across tasks and models.

\begin{enumerate}[leftmargin=*,nosep]
\item \textbf{Functional Extension}: add new functionality while preserving requirements established in earlier turns~\cite{lientz1978characteristics,lientz1980software,chapin2001types}.
\item \textbf{Error Handling Enhancement}: add validation and exception handling to existing code~\cite{miller1997exception}.
\item \textbf{Input Generalization}: extend the code suggestion to support additional input forms~\cite{lientz1980software,chapin2001types}.
\item \textbf{Performance Optimization}: improve efficiency while keeping outputs unchanged~\cite{arcelli2012performance}.
\item \textbf{Interface Restructuring}: refactor the program structure or interface while preserving requirements established in earlier turns~\cite{hosseini2008behavior}.
\item \textbf{Non-functional Augmentation}: add cross-cutting support such as logging or call statistics such as counts of calls, cache hits, and errors~\cite{kiczales1997aop}.
\end{enumerate}

All 542~tasks follow a \emph{fixed} 8-turn sequence with one baseline turn and seven controlled requirement changes (Table~\ref{tab:turn-sequence}). The ordering is designed so that later turns add new requirements while preserving those established earlier in the conversation, and the same sequence is applied to all tasks for comparability. Early turns establish the initial code suggestion and then add constraints through validation and broader input support. Turn~4 (T4) adds caching, Turn~5 (T5) adds new functionality, and Turn~6 (T6) refactors the accumulated code suggestion into a class. Turn~7 (T7) adds cross-cutting support such as logging or call statistics, and Turn~8 (T8) optimizes the class method's core algorithm while preserving caching, validation, type conversion, logging, and call statistics. The sequence also intentionally repeats some broad change families at different stages. Turn~5 (T5) revisits functional extension after validation, type generalization, and caching have already been added, and Turn~8 (T8) revisits algorithm optimization after the codebase has been restructured and augmented. This design lets us compare how similar modification families behave under different levels of accumulated conversational state. We do not claim that this is the only realistic ordering of maintenance turns.

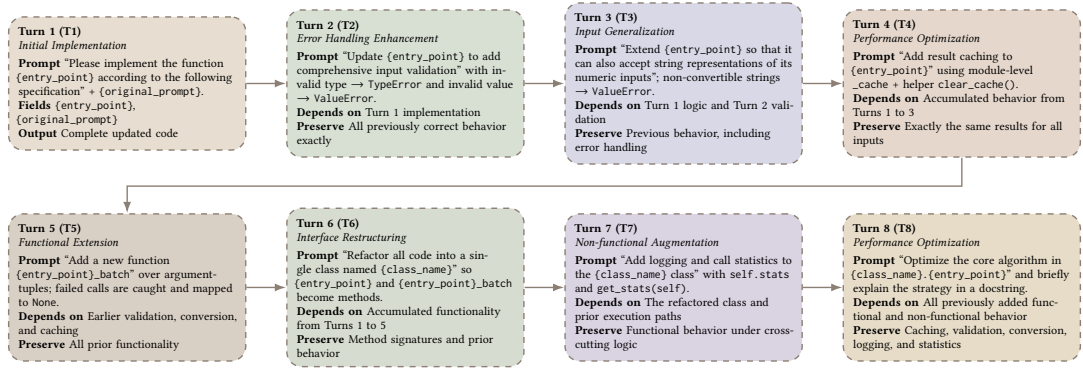
\begin{figure*}[!t]
\centering
\scriptsize

\definecolor{morandiA}{HTML}{E7DDD0}
\definecolor{morandiB}{HTML}{D8E0D4}
\definecolor{morandiC}{HTML}{D8D8E6}
\definecolor{morandiD}{HTML}{E5D7CC}
\definecolor{morandiE}{HTML}{D9D0C5}
\definecolor{morandiF}{HTML}{D6DCCF}
\definecolor{morandiG}{HTML}{D9D3E2}
\definecolor{morandiH}{HTML}{E6DCC8}
\definecolor{morandiLine}{HTML}{8C7F73}

\resizebox{0.8\textwidth}{!}{%
\begin{tikzpicture}[
  >=Latex,
  turn/.style n args={1}{
    draw=morandiLine,
    dashed,
    rounded corners=2.6mm,
    line width=0.7pt,
    fill=#1,
    align=left,
    text width=0.215\textwidth,
    inner sep=5pt,
    font=\scriptsize
  },
  flow/.style={
    -{Latex[length=2mm,width=1.4mm]},
    line width=0.7pt,
    draw=morandiLine
  }
]

\node[turn={morandiA}] (t1) {
\textbf{Turn 1 (T1)}\\
\textit{Initial Implementation}\\[2pt]
\textbf{Prompt} ``Please implement the function \texttt{\{entry\_point\}} according to the following specification'' + \texttt{\{original\_prompt\}}.\\
\textbf{Fields} \texttt{\{entry\_point\}}, \texttt{\{original\_prompt\}}\\
\textbf{Output} Complete updated code
};

\node[turn={morandiB}, right=7mm of t1] (t2) {
\textbf{Turn 2 (T2)}\\
\textit{Error Handling Enhancement}\\[2pt]
\textbf{Prompt} ``Update \texttt{\{entry\_point\}} to add comprehensive input validation'' with invalid type $\rightarrow$ \texttt{TypeError} and invalid value $\rightarrow$ \texttt{ValueError}.\\
\textbf{Depends on} Turn 1 implementation\\
\textbf{Preserve} All previously correct behavior exactly
};

\node[turn={morandiC}, right=7mm of t2] (t3) {
\textbf{Turn 3 (T3)}\\
\textit{Input Generalization}\\[2pt]
\textbf{Prompt} ``Extend \texttt{\{entry\_point\}} so that it can also accept string representations of its numeric inputs''; non-convertible strings $\rightarrow$ \texttt{ValueError}.\\
\textbf{Depends on} Turn 1 logic and Turn 2 validation\\
\textbf{Preserve} Previous behavior, including error handling
};

\node[turn={morandiD}, right=7mm of t3] (t4) {
\textbf{Turn 4 (T4)}\\
\textit{Performance Optimization}\\[2pt]
\textbf{Prompt} ``Add result caching to \texttt{\{entry\_point\}}'' using module-level \texttt{\_cache} + helper \texttt{clear\_cache()}.\\
\textbf{Depends on} Accumulated behavior from Turns 1 to 3\\
\textbf{Preserve} Exactly the same results for all inputs
};

\node[turn={morandiE}, below=11mm of t1] (t5) {
\textbf{Turn 5 (T5)}\\
\textit{Functional Extension}\\[2pt]
\textbf{Prompt} ``Add a new function \texttt{\{entry\_point\}\_batch}'' over argument-tuples; failed calls are caught and mapped to \texttt{None}.\\
\textbf{Depends on} Earlier validation, conversion, and caching\\
\textbf{Preserve} All prior functionality
};

\node[turn={morandiF}, right=7mm of t5] (t6) {
\textbf{Turn 6 (T6)}\\
\textit{Interface Restructuring}\\[2pt]
\textbf{Prompt} ``Refactor all code into a single class named \texttt{\{class\_name\}}'' so \texttt{\{entry\_point\}} and \texttt{\{entry\_point\}\_batch} become methods.\\
\textbf{Depends on} Accumulated functionality from Turns 1 to 5\\
\textbf{Preserve} Method signatures and prior behavior
};

\node[turn={morandiG}, right=7mm of t6] (t7) {
\textbf{Turn 7 (T7)}\\
\textit{Non-functional Augmentation}\\[2pt]
\textbf{Prompt} ``Add logging and call statistics to the \texttt{\{class\_name\}} class'' with \texttt{self.stats} and \texttt{get\_stats(self)}.\\
\textbf{Depends on} The refactored class and prior execution paths\\
\textbf{Preserve} Functional behavior under cross-cutting logic
};

\node[turn={morandiH}, right=7mm of t7] (t8) {
\textbf{Turn 8 (T8)}\\
\textit{Performance Optimization}\\[2pt]
\textbf{Prompt} ``Optimize the core algorithm in \texttt{\{class\_name\}.\{entry\_point\}}'' and briefly explain the strategy in a docstring.\\
\textbf{Depends on} All previously added functional and non-functional behavior\\
\textbf{Preserve} Caching, validation, conversion, logging, and statistics
};

\draw[flow] (t1.east) -- (t2.west);
\draw[flow] (t2.east) -- (t3.west);
\draw[flow] (t3.east) -- (t4.west);
\draw[flow] (t4.south) -- ($(t4.south)+(0,-5mm)$) -| (t5.north);
\draw[flow] (t5.east) -- (t6.west);
\draw[flow] (t6.east) -- (t7.west);
\draw[flow] (t7.east) -- (t8.west);

\end{tikzpicture}
}

\caption{Overview of the eight fixed turn templates used in the benchmark. For space reasons, the boxes show only the core content of each prompt rather than the full wording. Each turn uses fixed prompt wording, while benchmark-derived fields such as \texttt{\{entry\_point\}}, \texttt{\{original\_prompt\}}, and \texttt{\{class\_name\}} are filled deterministically. The arrows show the intended progression of the conversation from Turn 1 to Turn 8.}
\Description{A two-row overview of the eight turn templates used in the benchmark. Each box shows the turn number, turn type, a shortened prompt summary, the prior functionality it depends on, and the behavior it must preserve. Arrows connect the boxes from Turn 1 through Turn 8.}
\label{fig:turn-template-overview}
\end{figure*}

\smallskip
\noindent\textbf{Fixed Prompt Templates.}
We do not ask any LLM to invent a different follow-up turn for each task. Instead, for each turn we use the same fixed prompt wording across tasks and fill in only a small number of task-specific fields copied from the benchmark. These fields include the target function name (\texttt{entry\_point}), the benchmark's original task prompt at Turn~1 (T1), and the class name used for the refactoring turn at Turn~6 (T6), which is deterministically derived from \texttt{entry\_point}. Figure~\ref{fig:turn-template-overview} summarizes the eight turn templates. For example, the Turn~2 (T2) template always asks the model to add input validation while preserving requirements established in earlier turns. The only task-specific difference is the name of the function being modified. This design removes prompt variability as a confound while keeping the benchmark prompts recognizable and executable.

The templates provide a controlled and reproducible representation of common follow-up programming turns. We standardize the prompt structure and keep the wording stable, while allowing only benchmark-derived fields to vary. Some templates include concrete code hooks such as \texttt{\_cache}, \texttt{clear\_cache()}, and \texttt{get\_stats(self)}. These hooks make each turn-specific change operationally precise, comparable across tasks, and directly checkable by the same downstream regression tests.

As shown in Figure~\ref{fig:turn-template-overview}, \emph{Depends on} identifies which previously introduced functionality must be understood and modified to complete the current turn. \emph{Preserve} identifies which requirements established in earlier turns must remain intact after the change. Together, these two fields make the cross-turn dependency explicit. The resulting 8-turn sequence covers all six change types with restructuring at the midpoint, remains consistent with empirical evidence that developer-facing LLM interactions are typically short multi-turn conversations~\cite{hao2024shared}, and is long enough to expose regression accumulation across turns in the same conversation~\cite{laban2025llms}. In our setting, this length is also sufficient to reveal the full regression trajectory. The full wording of all eight templates is provided in our replication package (\url{https://anonymous.4open.science/r/multi-turn-llm-regression-E73E}).

All 542~tasks receive the same template at each turn. This choice ensures a controlled comparison across tasks, even though some turns may be less natural for certain problems and some prompt details may be more explicit than in unconstrained developer conversations. We accept this trade-off because the benchmark is designed to study regression under a fixed and reproducible sequence of evolving requirements. Our goal is not to replay the full diversity of natural maintenance conversations. It is intended to expose models to a stable set of follow-up changes that systematically require modifications of existing code while preserving requirements established in earlier turns. Our regression tests therefore evaluate whether later code suggestions preserve requirements established in earlier turns, not whether every turn is equally natural for every task.

\begin{table}[t]
\caption{Fixed 8-turn evolution sequence used for all 542 tasks.}\label{tab:turn-sequence}
\centering
\small
\begin{tabular}{@{}cll@{}}
\toprule
\textbf{Turn} & \textbf{Turn Type} & \textbf{Coupling Scope} \\
\midrule
T1  & Initial Code Suggestion & None (baseline) \\
T2  & Error Handling Enh.     & T1 control flow \\
T3  & Input Generalization    & T1 algorithm, T2 validation \\
T4  & Add Caching             & T1--T3 consistency \\
T5  & Functional Extension    & T1--T4 collectively \\
T6  & Interface Restructuring & T1--T5 all requirements \\
T7  & Non-functional Aug.     & T2,T4,T6 cross-cutting \\
T8  & Algorithm Opt.          & T1--T7 full codebase \\
\bottomrule
\end{tabular}
\end{table}

\subsection{Taxonomy Development}\label{sec:taxonomy}

To characterize failures that emerge only after requirements accumulate across turns, we develop a new taxonomy of multi-turn bugs and use it as the coding scheme for the annotation study. This taxonomy is not directly adopted from an existing bug taxonomy. Instead, it is derived bottom-up from our failure cases while being informed by qualitative coding and thematic-synthesis practice in software engineering~\cite{seaman1999qualitative,cruzes2011thematic,stol2016grounded}. We began with an initial calibration subset of 100 failure cases, selected to cover turn positions, change types, models, and failure severities, and asked the three coders (all co-authors) to independently read and open-code the observed failure patterns without pre-fixing the final labels. This bottom-up process produced a provisional pool of fine-grained failure codes rather than predefined high-level classes. The three coders then compared their codes, merged synonymous or overlapping concepts, split labels whose boundary covered more than one mechanism, and removed candidate codes that were too rare, too ambiguous, or not distinguishable enough to support stable annotation rules. We refined the remaining labels by writing explicit definitions, inclusion criteria, exclusion criteria, and illustrative examples for each candidate label. A label was retained in the final scheme only if it captured a recurring failure pattern, had a boundary that annotators could distinguish from neighboring labels, and could be supported by a concrete decision rule.

Across subsequent calibration rounds, the same three authors applied the evolving codebook to shared cases and used disagreements to refine the label set rather than to optimize agreement prematurely. These disagreements repeatedly traced back to three issues: label overlap, overly abstract definitions, and underspecified decision rules. We therefore added distinguishing rules and made the single-label decision rule explicit for boundary cases.

Once the concrete labels and their boundaries stabilized, we grouped them into higher-level classes based on their primary root cause. In other words, this higher-level organization was introduced after the label set had stabilized, not imposed before coding. We aligned this organization with established software engineering concepts such as regression bugs~\cite{yoo2012regression}, code decay~\cite{belady1976model}, and feature interaction~\cite{zave1993feature}. We also compared the resulting scheme against existing bug taxonomies for AI-generated code suggestions~\cite{abbassi2025taxonomy} to identify multi-turn failure patterns that are not well captured by prior single-turn classifications. After the label set and class structure stabilized, we froze the codebook for the full annotation study.

The finalized scheme operationalizes annotation as a bug-classification form rather than a free-response questionnaire. After freezing the codebook, four of the study's authors independently applied it to the 384 sampled failures. For each failed turn instance, annotators inspect the original task prompt, the turn number and change type, the current-turn prompt for the failing turn (for example, Turn~$n$ or T$n$), the last correct code suggestion from the previous successful turn (for example, Turn~$n\!-\!1$), the failing code suggestion, the failed tests and error messages, and the regression rate. They then diagnose which previously established requirement or constraint has been broken and encode that diagnosis by assigning one bug label from the frozen codebook. We present these options to annotators as closed-choice labels grouped under the higher-level classes, with Baseline Failure reserved for Turn~1 cases. The high-level failure mechanism is inferred from the selected label's class membership rather than being recorded through a separate question. After full annotation begins, the taxonomy is no longer substantially revised, so that agreement statistics remain interpretable. We apply the frozen codebook in the 384-sample annotation study described in Section~\ref{sec:annotation-protocol}. We present the taxonomy itself, its agreement statistics, and its empirical distribution in \RQtwo{}.

\subsection{Mitigation Strategy Design}\label{sec:mitigation-approach}

We design mitigation strategies for two risks that arise in multi-turn coding conversations. One is that the model may lose track of constraints established earlier in the conversation, which is consistent with prior observations of multi-turn quality loss and context drift~\cite{laban2025llms,dongre2025driftnomore}. The other is that a new code suggestion may break requirements established in earlier turns when a fresh requirement is applied to already-evolving code, a recurring concern in change-impact and ripple-effect analysis~\cite{yau1978ripple,bohner2002impacts}. We therefore formulate two lightweight interaction-level strategies and evaluate them individually and in combination. Both strategies operate within the ongoing conversation rather than as post-hoc repair procedures, allowing us to assess whether they reduce regression as requirements accumulate.

\subsubsection{S1: Snowball Recap.}
Snowball Recap is designed for cases in which the model may lose track of requirements established earlier in the conversation. In our protocol, each new turn is appended to the existing dialogue, so later turns are issued under an increasingly long interaction history. The strategy therefore makes earlier constraints explicit again before each later turn instead of relying on the model to recover them from the growing conversation context alone. Following prior evidence that restating relevant context can reduce quality loss in long LLM interactions~\cite{laban2025llms} and prior observations that long conversations can cause models to lose track of earlier requirements~\cite{dongre2025driftnomore}, this strategy prepends a recap prefix before each turn $t \geq 2$. The prefix has three fixed parts: a header stating that all requirements from earlier turns must continue to hold, one summary line for each earlier turn in the format ``T$i$. [change type]: summary.'', and a fixed transition line before the prompt for the current turn. These summary lines are not generated by another model. Instead, they are produced deterministically from two metadata fields already stored in the benchmark instance, namely the turn index and the turn's change type. For example, Turn~2 contributes the recap line ``T2. [Error Handling Enhancement]: add input validation (TypeError/ValueError) to all public methods without rejecting previously valid inputs,'' and Turn~3 contributes ``T3. [Input Generalization]: accept string representations of numeric inputs and convert them, while preserving existing validation and logic.'' For Turn~4, the final prompt is the fixed header, followed by the recap lines for Turns~2 and~3, and then the original Turn~4 prompt.

\subsubsection{S2: Verification Gate.}
Verification Gate is designed for cases in which a newly generated code suggestion immediately breaks requirements that were preserved at the previous turn, a familiar regression risk under successive change~\cite{yoo2012regression,yau1978ripple,bohner2002impacts}. Unlike post-hoc repair methods that ask the model to fix a code suggestion only after it has already failed~\cite{chen2024selfdebug,kumar2025score}, this strategy checks each turn immediately after generation and applies the one-step recoverability principle of Jain et al.~\cite{jain2025mucode}. After each turn $t \geq 2$, we execute the same benchmark-provided tests used throughout the study and compare the current regression pass rate with that of the preceding turn. The gate is triggered only when $RPR_t < RPR_{t-1}$ and a last passing code suggestion is available. In that case, the current code suggestion is rejected and the model receives a rollback prompt with three fixed elements: the number of failed regression tests, the full code of the last passing code suggestion, and the original requirement for the current turn. Concretely, the prompt tells the model that its previous attempt caused $n_{\text{fail}}$ of $n_{\text{total}}$ regression tests to fail, provides the last correct code block, and asks it to apply the current requirement to that version instead. The model is then given a single retry. We retain the retry only when it passes at least as many benchmark tests as the rejected code suggestion; otherwise, the current code suggestion is kept. For example, suppose Turn~3 has already added support for string representations of numeric inputs, and Turn~4 asks the model to add caching. If the first Turn~4 code suggestion adds caching but breaks the earlier input-conversion requirement, the gate rejects that code suggestion, restores the last passing Turn~3 version, and asks the model to generate Turn~4 again from that point. This design bounds intervention cost to at most one additional generation per turn while making the rollback criterion fully observable and reproducible.

\subsubsection{S3: Combined (S1\,+\,S2).}
The combined strategy applies both recap and verification gating in the same conversation. It tests whether the two interventions address complementary failure mechanisms or largely overlap in practice.

All three strategies are model-agnostic. They require neither fine-tuning nor access to the internal model. Their overhead is also bounded. Snowball Recap adds prompt tokens only, and Verification Gate introduces at most one retry per turn.

\section{Experimental Setup}\label{sec:setup}

\subsection{Models Under Study}

We evaluate six LLMs, including two commercial models (GPT-4o~\cite{openai2024gpt4o} and DeepSeek-V3~\cite{deepseek2024v3}) and four open-source models (Qwen2.5-Coder-32B~\cite{qwen2024qwen25coder}, Qwen3-32B~\cite{qwen2025qwen3}, DS-R1-Distill-32B~\cite{deepseek2025r1}, and Llama-3.1-8B~\cite{meta2024llama31}). This set allows us to compare commercial models and to compare a code-specialized model (i.e., Qwen2.5-Coder-32B) with general-purpose models (i.e., GPT-4o, DeepSeek-V3, Qwen3-32B, and Llama-3.1-8B). It also includes a reasoning-oriented model (i.e., DS-R1-Distill-32B) alongside standard-generation models.% and a smaller local model (i.e., Llama-3.1-8B) with larger local models (i.e., the 32B open-source models).

The commercial models are accessed through APIs, whereas the open-source models are evaluated under a shared local inference setup. Within the local group, Qwen2.5-Coder-32B and Qwen3-32B isolate code specialization at the same nominal size, DS-R1-Distill-32B adds a reasoning-oriented variant, and Llama-3.1-8B provides a smaller scale contrast.

All four open-source models are served locally on a single 32\,GB NVIDIA RTX 5090 through vLLM~\cite{kwon2023efficient} with AWQ 4-bit quantization and \texttt{max-model-len} set to 32,768. GPT-4o and DeepSeek-V3 are accessed through their official OpenAI-compatible APIs.

\begin{table}[t]
\caption{Models evaluated in the study.}\label{tab:models}
\centering
\small
\resizebox{\linewidth}{!}{\begin{tabular}{@{}lllll@{}}
\toprule
\textbf{Model} & \textbf{Params} & \textbf{Type} & \textbf{Deploy} & \textbf{Comparison} \\
\midrule
GPT-4o             & ---   & General  & API   & Commercial baseline \\
DeepSeek-V3        & 671B  & General  & API   & MoE architecture \\
\midrule
Qwen2.5-Coder-32B  & 32B   & Code     & Local & Code-specialized \\
Qwen3-32B           & 32B   & General  & Local & General (same family) \\
DS-R1-Distill-32B   & 32B   & Reason   & Local & Reasoning model \\
Llama-3.1-8B        & 8B    & General  & Local & Scale comparison \\
\bottomrule
\end{tabular}}
\end{table}

\subsection{Evaluation Protocol}

For each task on each model, we run one 8-turn conversation consisting of an initial code-suggestion turn followed by seven controlled requirement changes. All six models are evaluated on the same 542 tasks, comprising 164 HumanEval+ tasks and 378 MBPP+ tasks, totaling 3,252 conversations and 26,016 turn instances.

Each conversation starts with a fixed system prompt that asks the model to return the full code suggestion in a single Python code block. At Turn~1, the user message is the benchmark prompt for the initial code suggestion. At each subsequent turn, the user message is the fixed benchmark request for that turn. At Turn~$t$, the input contains the system prompt, the user requests from Turns~1 to~$t$, and the model responses from all earlier turns.

After each response, we extract the code suggestion by taking the last fenced Python block. If no fenced block is present, we use the raw response text as the candidate program. We then evaluate that program against the regression oracle for the current turn. For function-level tasks, Turns~1 to~5 use the benchmark's original tests, and Turns~6 to~8 use deterministically adapted versions of those same tests after the interface is restructured into a class. The adaptation prepends object instantiation and rewrites direct function calls into method calls, but leaves the original assertions unchanged. For each test, we build a temporary Python script containing the code suggestion and the test, run it in an isolated subprocess, and enforce a 15-second timeout. Assert-style tests are executed independently and in parallel, whereas unittest-style suites are executed in a single subprocess.

\subsection{Metrics}\label{sec:metrics}

We use one turn-level metric, Regression Pass Rate, which captures how much previously correct behavior is still preserved at each turn. The remaining three metrics summarize regression accumulation over the full conversation.

\begin{itemize}[leftmargin=*,nosep]
\item \textbf{Regression Pass Rate} ($RPR_t$). For a given task at turn $t$, $RPR_t = p_t / n_t$, where $p_t$ is the number of benchmark tests that pass at turn $t$ and $n_t$ is the total number of benchmark tests for that task. Because the same benchmark tests are preserved across turns, a lower $RPR_t$ directly indicates that more prior behavior has been broken.
\item \textbf{Degradation} ($\Delta$). $\Delta = RPR_8 - RPR_1$. This metric captures the net change in quality from the first to the last turn.
\item \textbf{Degradation Rate}. This is the fraction of tasks for which $\Delta < 0$. It measures how often quality declines over the conversation.
\item \textbf{Survival Rate}. This metric is computed only for tasks with $RPR_1 = 100\%$. It reports the fraction of those initially solved tasks that still satisfy $RPR_t \geq 50\%$ at turn $t$.
\end{itemize}

\subsection{Bug Annotation Protocol}\label{sec:annotation-protocol}

To answer \RQtwo{}, we annotate a stratified sample of failure instances. We set the sample size to 384 using Cochran's standard formula for proportion estimates with a 95\% confidence level, a 5\% margin of error, and the conservative choice $p=0.5$~\cite{cochran1977sampling}. From all turn instances with $RPR_t < 1.0$, we then stratify this sample across turn number and model so that the annotation set covers the main parts of the experimental space.

The four annotators are the first four authors. Each of the four independently labels every sampled case using the frozen codebook developed in Section~\ref{sec:taxonomy}. For each sample, annotators inspect the most recent earlier code suggestion that passed the regression oracle, the failing code suggestion at the current turn, the failed tests and error messages, and the current-turn request. They then assign one bug label from the predefined codebook as the root-cause label.

To reduce within-annotator inconsistency, each annotator labels every sample in three passes and keeps the within-annotator majority label as the final individual judgment. We then aggregate the four final judgments across annotators. Of the 384 samples, 277 are unanimous, and 107 are non-unanimous. For non-unanimous cases with a majority label, we keep the majority label. For the remaining 42 cases without a majority label, the senior researcher resolves the tie. We report inter-rater agreement in \RQtwo{} using Fleiss' $\kappa$, because each of the 384 samples receives one nominal bug label from each of four annotators, and Fleiss' $\kappa$ provides a chance-corrected agreement measure for more than two raters, as in prior software engineering studies with multi-rater labeling~\cite{bambazek2023requirements,gonzalez2023reliability}. We compute it on the final annotator labels before majority aggregation so that it reflects how consistently the bug taxonomy can be applied independently.

\section{Results}\label{sec:results}

% =============================================================
\subsection{\RQone{}: \RQoneTitle{}}\label{sec:rq1}
% =============================================================

\begin{figure}[t]
\centering
\resizebox{0.88\columnwidth}{!}{%
\begin{tikzpicture}[x=0.74cm,y=0.09cm]
  \definecolor{gridColor}{HTML}{D9DEE3}
  \definecolor{textColor}{HTML}{2F343B}
  \definecolor{gptColor}{HTML}{486F97}
  \definecolor{deepseekColor}{HTML}{4F9E9B}
  \definecolor{qwenCoderColor}{HTML}{BC7A56}
  \definecolor{qwen3Color}{HTML}{7A9950}
  \definecolor{dsr1Color}{HTML}{66717F}
  \definecolor{llamaColor}{HTML}{B86764}

  % Axes and grid
  \foreach \y in {20,30,40,50,60,70,80,90} {
    \draw[color=gridColor, line width=0.3pt] (1,\y) -- (8,\y);
    \node[anchor=east, font=\scriptsize, text=textColor] at (0.82,\y) {\y};
  }
  \draw[line width=0.5pt, color=black!60] (1,20) -- (8.2,20);
  \draw[line width=0.5pt, color=black!60] (1,20) -- (1,92);

  \foreach \x in {1,2,3,4,5,6,7,8} {
    \draw[color=black!60] (\x,20) -- (\x,21.2);
    \node[anchor=north, font=\scriptsize, text=textColor] at (\x,18.3) {T\x};
  }

  \node[font=\scriptsize, text=textColor] at (4.5,14.4) {Conversation turn};
  \node[rotate=90, font=\scriptsize, text=textColor] at (0.18,56) {Regression Pass Rate (\%)};

  % GPT-4o
  \draw[gptColor, line width=0.95pt, rounded corners=2.2pt]
    (1,90.2) -- (2,79.9) -- (3,75.9) -- (4,75.1) -- (5,75.1) -- (6,74.9) -- (7,74.8) -- (8,72.4);
  \foreach \x/\y in {1/90.2,2/79.9,3/75.9,4/75.1,5/75.1,6/74.9,7/74.8,8/72.4}
    \fill[gptColor] (\x,\y) circle (0.07);

  % DeepSeek-V3
  \draw[deepseekColor, line width=0.95pt, rounded corners=2.2pt]
    (1,87.6) -- (2,81.3) -- (3,80.4) -- (4,78.4) -- (5,78.5) -- (6,78.5) -- (7,78.3) -- (8,75.8);
  \foreach \x/\y in {1/87.6,2/81.3,3/80.4,4/78.4,5/78.5,6/78.5,7/78.3,8/75.8}
    \draw[deepseekColor, line width=0.9pt] (\x-0.07,\y-0.7) rectangle (\x+0.07,\y+0.7);

  % Qwen2.5-Coder-32B
  \draw[qwenCoderColor, line width=0.95pt, dash pattern=on 3pt off 1.5pt, rounded corners=2.2pt]
    (1,84.5) -- (2,75.4) -- (3,63.5) -- (4,62.4) -- (5,62.3) -- (6,61.8) -- (7,58.8) -- (8,55.0);
  \foreach \x/\y in {1/84.5,2/75.4,3/63.5,4/62.4,5/62.3,6/61.8,7/58.8,8/55.0}
    \fill[qwenCoderColor] (\x,\y+0.82) -- (\x-0.08,\y-0.62) -- (\x+0.08,\y-0.62) -- cycle;

  % Qwen3-32B
  \draw[qwen3Color, line width=0.95pt, dash pattern=on 1.2pt off 1.2pt, rounded corners=2.2pt]
    (1,72.7) -- (2,62.8) -- (3,49.5) -- (4,59.3) -- (5,63.8) -- (6,66.5) -- (7,59.7) -- (8,57.9);
  \foreach \x/\y in {1/72.7,2/62.8,3/49.5,4/59.3,5/63.8,6/66.5,7/59.7,8/57.9} {
    \draw[qwen3Color, line width=0.8pt] (\x,\y+0.9) -- (\x-0.08,\y) -- (\x,\y-0.9) -- (\x+0.08,\y) -- cycle;
  }

  % DS-R1-Distill-32B
  \draw[dsr1Color, line width=0.95pt, dash pattern=on 5pt off 1.5pt on 1pt off 1.5pt, rounded corners=2.2pt]
    (1,77.5) -- (2,71.3) -- (3,24.7) -- (4,47.6) -- (5,51.4) -- (6,56.8) -- (7,49.9) -- (8,51.7);
  \foreach \x/\y in {1/77.5,2/71.3,3/24.7,4/47.6,5/51.4,6/56.8,7/49.9,8/51.7} {
    \draw[dsr1Color, line width=0.8pt] (\x-0.08,\y-0.8) -- (\x+0.08,\y+0.8);
    \draw[dsr1Color, line width=0.8pt] (\x-0.08,\y+0.8) -- (\x+0.08,\y-0.8);
  }

  % Llama-3.1-8B
  \draw[llamaColor, line width=0.95pt, dash pattern=on 2pt off 1.2pt, rounded corners=2.2pt]
    (1,76.1) -- (2,63.0) -- (3,49.5) -- (4,47.8) -- (5,47.8) -- (6,47.4) -- (7,47.7) -- (8,31.6);
  \foreach \x/\y in {1/76.1,2/63.0,3/49.5,4/47.8,5/47.8,6/47.4,7/47.7,8/31.6} {
    \draw[llamaColor, line width=0.8pt] (\x-0.08,\y) -- (\x+0.08,\y);
    \draw[llamaColor, line width=0.8pt] (\x,\y-0.8) -- (\x,\y+0.8);
  }

  % Legend
  \draw[gptColor, line width=0.95pt] (0.95,97.0) -- (1.25,97.0);
  \fill[gptColor] (1.10,97.0) circle (0.07);
  \node[anchor=west, font=\scriptsize, text=textColor] at (1.45,97.0) {GPT-4o};

  \draw[deepseekColor, line width=0.95pt] (3.05,97.0) -- (3.35,97.0);
  \draw[deepseekColor, line width=0.9pt] (3.20-0.07,97.0-0.7) rectangle (3.20+0.07,97.0+0.7);
  \node[anchor=west, font=\scriptsize, text=textColor] at (3.55,97.0) {DeepSeek-V3};

  \draw[qwenCoderColor, line width=0.95pt, dash pattern=on 3pt off 1.5pt] (5.55,97.0) -- (5.85,97.0);
  \fill[qwenCoderColor] (5.70,97.82) -- (5.62,96.38) -- (5.78,96.38) -- cycle;
  \node[anchor=west, font=\scriptsize, text=textColor] at (6.10,97.0) {Qwen2.5-Coder-32B};

  \draw[qwen3Color, line width=0.95pt, dash pattern=on 1.2pt off 1.2pt] (0.95,100.4) -- (1.25,100.4);
  \draw[qwen3Color, line width=0.8pt] (1.10,101.3) -- (1.02,100.4) -- (1.10,99.5) -- (1.18,100.4) -- cycle;
  \node[anchor=west, font=\scriptsize, text=textColor] at (1.45,100.4) {Qwen3-32B};

  \draw[dsr1Color, line width=0.95pt, dash pattern=on 5pt off 1.5pt on 1pt off 1.5pt] (3.05,100.4) -- (3.35,100.4);
  \draw[dsr1Color, line width=0.8pt] (3.20-0.08,100.4-0.8) -- (3.20+0.08,100.4+0.8);
  \draw[dsr1Color, line width=0.8pt] (3.20-0.08,100.4+0.8) -- (3.20+0.08,100.4-0.8);
  \node[anchor=west, font=\scriptsize, text=textColor] at (3.55,100.4) {DS-R1-Distill-32B};

  \draw[llamaColor, line width=0.95pt, dash pattern=on 2pt off 1.2pt] (6.55,100.4) -- (6.85,100.4);
  \draw[llamaColor, line width=0.8pt] (6.70-0.08,100.4) -- (6.70+0.08,100.4);
  \draw[llamaColor, line width=0.8pt] (6.70,100.4-0.8) -- (6.70,100.4+0.8);
  \node[anchor=west, font=\scriptsize, text=textColor] at (7.05,100.4) {Llama-3.1-8B};
\end{tikzpicture}
}
\caption{Regression pass rate trajectories across the eight-turn conversation for all six models ($N$\,=\,542 for each model). All models end below their Turn~1 pass rate, although the timing and severity of the losses differ by model.}
\Description{A line chart showing regression pass rate from Turn 1 to Turn 8 for six models. DeepSeek-V3 and GPT-4o remain highest across the conversation. Llama-3.1-8B drops the most. The other 32B open models fall in between, with notable early drops around Turns 2 and 3.}
\label{fig:summary-all-models}
\end{figure}
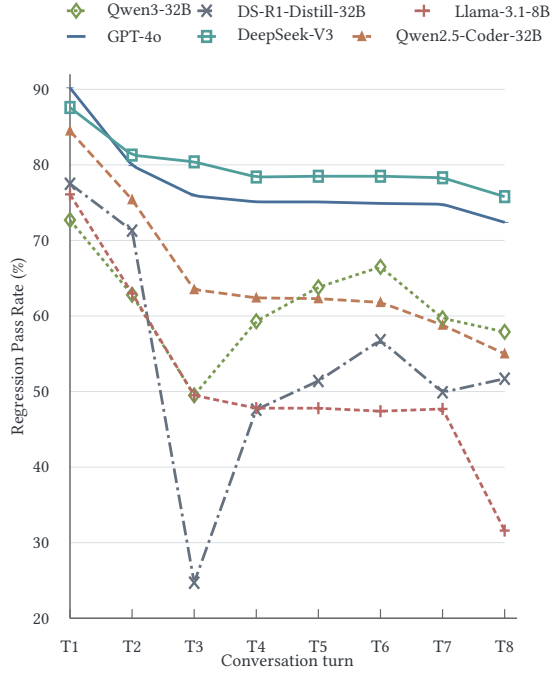

%\noindent 
We organize the results in three parts. Figure~\ref{fig:summary-all-models} shows the end-to-end trajectories from Turn~1 to Turn~8 for all six models. We report per-turn $RPR$ values and identify where the largest losses occur in Table~\ref{tab:per-turn-all}. Table~\ref{tab:initially-solved} shows the same pattern after restricting each model to tasks it solved at Turn~1.

\noindent\textbf{Finding 1. Regression accumulates for all six models, but its magnitude differs substantially by model family.} As shown in Figure~\ref{fig:summary-all-models}, every trajectory trends downward from Turn~1 to Turn~8. No model finishes at or above its initial regression pass rate. Across models, 40--73\% of tasks lose previously correct behavior over the full conversation. DeepSeek-V3 and GPT-4o remain strongest at Turn~8, reaching 75.8\% and 72.4\% respectively. The 32B open models form a middle tier, with final-turn $RPR$ values ranging from 51.7\% to 57.9\%. Llama-3.1-8B shows the largest loss and falls to 31.6\%.

\noindent\textbf{Finding 2. Strong Turn~1 performance does not necessarily translate into stronger multi-turn resilience.} DeepSeek-V3 begins below GPT-4o at Turn~1, yet retains more quality by Turn~8. Llama-3.1-8B declines much more sharply than the larger commercial models and finishes far below the 32B open models at Turn~8. Reasoning specialization does not appear to help in this setting, since DS-R1-Distill-32B reaches 51.7\% at Turn~8, whereas Qwen3-32B reaches 57.9\%.

\noindent\textbf{Finding 3. The per-turn results show that regression losses are front-loaded and are concentrated in a small number of change types.} For five of the six models, the largest losses occur at Turn~2 or Turn~3. These error-handling and input-generalization turns are disruptive because the new validation or conversion logic must remain compatible with both the task's original behavior and earlier-turn requirements.

\begin{table*}[t]
\caption{Per-turn $RPR$ and LOC for all six models ($N$\,=\,542). Bold marks the largest per-turn $RPR$ loss for each model.}\label{tab:per-turn-all}
\centering
\scriptsize
\setlength{\tabcolsep}{3pt}
\resizebox{0.76\textwidth}{!}{%
\begin{tabular}{@{}cl|rr|rr|rr|rr|rr|rr@{}}
\toprule
& & \multicolumn{2}{c|}{\textbf{GPT-4o}} & \multicolumn{2}{c|}{\textbf{DeepSeek}} & \multicolumn{2}{c|}{\textbf{Qwen-Coder}} & \multicolumn{2}{c|}{\textbf{Qwen3}} & \multicolumn{2}{c|}{\textbf{DS-R1}} & \multicolumn{2}{c}{\textbf{Llama-8B}} \\
\textbf{T} & \textbf{Turn Type} & $RPR$ & LOC & $RPR$ & LOC & $RPR$ & LOC & $RPR$ & LOC & $RPR$ & LOC & $RPR$ & LOC \\
\midrule
1 & Initial Impl.     & 90.2 &  9 & 87.6 & 15 & 84.5 & 18 & 72.7 & 50 & 77.5 & 43 & 76.1 &  8 \\
2 & Error Handling     & 79.9 & 18 & 81.3 & 31 & 75.4 & 30 & 62.8 & 76 & 71.3 & 47 & \textbf{63.0} & 16 \\
3 & Input General.     & \textbf{75.9} & 28 & 80.4 & 49 & \textbf{63.5} & 42 & \textbf{49.5} & 99 & \textbf{24.7} &175 & 49.5 & 25 \\
4 & Add Caching       & 75.1 & 42 & 78.4 & 74 & 62.4 & 59 & 59.3 & 88 & 47.6 &124 & 47.8 & 37 \\
5 & Func.\ Ext.        & 75.1 & 53 & 78.5 &103 & 62.3 & 79 & 63.8 & 94 & 51.4 &124 & 47.8 & 49 \\
6 & Interface Restr.   & 74.9 & 55 & 78.5 &107 & 61.8 & 83 & 66.5 & 89 & 56.8 &122 & 47.4 & 50 \\
7 & Non-func.\ Aug.    & 74.8 & 79 & 78.3 &147 & 58.8 & 94 & 59.7 &135 & 49.9 &163 & 47.7 & 67 \\
8 & Algorithm Opt.    & 72.4 & 85 & \textbf{75.8} &163 & 55.0 & 87 & 57.9 &139 & 51.7 &166 & 31.6 & 71 \\
\bottomrule
\end{tabular}}% end resizebox
\setlength{\tabcolsep}{6pt}
\end{table*}

No model shows sustained recovery once quality drops. Some local improvements appear, such as Qwen3 between T5 and T6, but they do not reverse the overall trend. In other words, once regression appears in the conversation, later turns rarely restore the lost behavior on their own.

\noindent\textbf{Finding 4. Regression accumulation appears for both easier and harder seed tasks.} For DeepSeek-V3, the hard group ($\geq$9~LOC in the benchmark reference solution) declines from 82.5\% at Turn~1 to 62.7\% at Turn~8, while the easy group (1 to 3~LOC) declines from 88.3\% to 77.4\%.

\noindent\textbf{Finding 5. Regression accumulation remains substantial even among tasks solved perfectly at Turn~1 (\textit{i.e.,} tasks with $RPR_{T1}=100\%$).} To isolate the regression signal from baseline inability, we perform this analysis separately for each model and retain only the tasks on which that model passes all benchmark tests at Turn~1. We then report the median task-level $RPR_{T8}$, the percentage of tasks with $RPR_{T8}<100\%$, and the percentage of tasks with $RPR_{T8}=0\%$ in Table~\ref{tab:initially-solved}.

\begin{table}[t]
\caption{Regression outcomes among initially solved tasks ($RPR_{T1}=100\%$). ``Med.'' denotes the median final-turn regression pass rate, and ``Collapsed'' denotes $RPR_{T8}=0\%$.}\label{tab:initially-solved}
\centering
\small
\begin{tabular}{@{}lrrrr@{}}
\toprule
\textbf{Model} & \textbf{$N$} & \textbf{Med.\ $RPR_{T8}$} & \textbf{\% With Regr.} & \textbf{\% Collapsed} \\
\midrule
GPT-4o            & 427 & 100.0\% & 47.8\% &  3.0\% \\
DeepSeek-V3       & 414 & 100.0\% & 45.9\% &  1.9\% \\
Qwen2.5-Coder-32B & 440 & 98.1\% & 54.5\% & 14.3\% \\
Qwen3-32B          & 397 & 98.2\% & 50.9\% & 10.6\% \\
DS-R1-Distill-32B  & 428 & 92.8\% & 61.2\% & 29.2\% \\
Llama-3.1-8B       & 333 & 16.1\% & 81.7\% & 34.2\% \\
\bottomrule
\end{tabular}
\end{table}

The ordering of the models by median task-level $RPR_{T8}$ is unchanged after restricting each model to tasks it solved at T1. On this restricted subset, the median task-level $RPR_{T8}$ remains 100\% for GPT-4o and DeepSeek-V3, while the 32B open models also retain high median $RPR_{T8}$ values but still exhibit collapse rates ranging from 10.6\% to 29.2\%. The two weakest models, i.e., DS-R1-Distill-32B and Llama-3.1-8B, lose a substantial fraction of the tasks they solve at Turn~1 (T1) to total failure by Turn~8.

\begin{tcolorbox}[colback=black!2,colframe=black!20,boxrule=0.4pt,arc=3pt,left=4pt,right=4pt,top=4pt,bottom=4pt]
\small
Regression accumulates for all six LLM models in multi-turn conversations and are concentrated in specific change types (i.e., error-handling and input-generalization requests).
Subsequent changes to requirements in a multi-turn conversation can introduce regression even when the first-turn code suggestion is correct.
\end{tcolorbox}

% =============================================================
\subsection{\RQtwo{}: \RQtwoTitle{}}\label{sec:rq2}
% =============================================================

\noindent To answer RQ2, we analyze 384 manually labeled failure cases (Section~\ref{sec:annotation-protocol}). Table~\ref{tab:bug-taxonomy-dist} reports the taxonomy and its distribution. 
The reported counts and percentages use the finalized label per case, so each annotated case contributes to exactly one row in the distribution. Baseline Failure (BF) is reserved for Turn~1 failure and is therefore interpreted separately from multi-turn regression failures.

\begin{table}[t]
\caption{Multi-turn bug taxonomy and bug type distribution from four-annotator labels (384 samples). $\star$ marks types not well represented in prior single-turn taxonomies. BF applies only to T1 failures.}\label{tab:bug-taxonomy-dist}
\centering
\scriptsize
\setlength{\tabcolsep}{2.5pt}
\resizebox{\columnwidth}{!}{%
\begin{tabular}{@{}llp{3.35cm}rr@{}}
\toprule
\textbf{ID} & \textbf{Name} & \textbf{Definition} & \textbf{Count} & \textbf{\%} \\
\midrule
\multicolumn{5}{@{}l}{\textit{Baseline}} \\
BF & Baseline Failure & T1 implementation error, not a multi-turn bug & 48 & 12.5 \\
\midrule
\multicolumn{5}{@{}l}{\textit{Class I: Context-Loss}} \\
CL-1 & Context Loss & Forgets or overwrites a constraint from a prior turn & 63 & 16.4 \\
CL-3 & Partial Recall & Remembers function structure but misses internal details & 18 & 4.7 \\
\multicolumn{3}{@{}r}{\textit{Subtotal}} & \textit{81} & \textit{21.1} \\
\midrule
\multicolumn{5}{@{}l}{\textit{Class II: Cross-Turn Conflict}} \\
CC-1 & Semantic Collision & New logic contradicts prior logic, producing wrong results & 102 & 26.6 \\
CC-2 & Interface Mismatch & New code breaks prior interfaces or signatures & 1 & 0.3 \\
CC-3 & Integration Inconsistency & Individually correct components fail to integrate & 1 & 0.3 \\
CC-4 & Over-Guarding & New validation rejects inputs the specification treats as valid & 110 & 28.6 \\
\multicolumn{3}{@{}r}{\textit{Subtotal}} & \textit{214} & \textit{55.7} \\
\midrule
\multicolumn{5}{@{}l}{\textit{Class III: Accumulation}} \\
AC-1$^\star$ & Regression via Omission & Silently drops necessary prior code during regeneration & 3 & 0.8 \\
AC-2$^\star$ & Complexity Collapse & Code exceeds capacity, causing syntax errors or truncation & 38 & 9.9 \\
\multicolumn{3}{@{}r}{\textit{Subtotal}} & \textit{41} & \textit{10.7} \\
\bottomrule
\end{tabular}}
\setlength{\tabcolsep}{6pt}
\end{table}

\noindent\textbf{Finding 6. Cross-Turn Conflict is the major source of manually observed regression failures.} As shown in Table~\ref{tab:bug-taxonomy-dist}, Cross-Turn Conflict accounts for 55.7\% of labeled cases, compared with 21.1\% for Context-Loss and 10.7\% for Accumulation. This indicates that later revisions more often break previously correct behavior through incompatibility with existing code than through simple forgetting.

\noindent\textbf{Finding 7. The distribution is concentrated in two conflict types, and their locations across change requests are interpretable.} Over-Guarding (CC-4) contributes 28.6\% and Semantic Collision (CC-1) contributes 26.6\% of all labeled failures. Together they account for 55.2\% of all annotated cases. Within CC-4, more than half of the cases arise from error-handling and input-generalization requests (59/110, 53.6\%), in which stricter checks must preserve previously valid behavior. Within CC-1, half of the cases involve functional extension and performance optimization (51/102, 50.0\%), in which new logic is added on top of existing behavior.

\noindent\textbf{Finding 8. Accumulation-specific failures are present, but less frequent.} Complexity Collapse (AC-2) appears in 9.9\% of annotated failures and Regression via Omission (AC-1) appears in 0.8\%. These patterns are explicitly separated in our taxonomy because they are not well represented in prior single-turn LLM bug taxonomies~\cite{abbassi2025taxonomy}. Four annotators independently labeled all 384 samples, with unanimous agreement on 277 samples and Fleiss' $\kappa=0.799$ (\emph{substantial} agreement~\cite{landis1977measurement}), which supports the reliability of the reported distribution.

\begin{tcolorbox}[colback=black!2,colframe=black!20,boxrule=0.4pt,arc=3pt,left=4pt,right=4pt,top=4pt,bottom=4pt]
\small
Manual labels indicate that multi-turn regression is primarily an integration problem rather than solely a Context-Loss problem.
Over-Guarding and Semantic Collision account for most cases, while accumulation-specific failures remain observable as code grows.
\end{tcolorbox}

\subsection{\RQthree{}: \RQthreeTitle{}}\label{sec:rq3-mitigation}
% =============================================================

\noindent To answer RQ3, we evaluate whether interaction-level controls can mitigate regression accumulation during the conversation itself. We compare Snowball Recap (S1), Verification Gate (S2), and their combination (S3). S1 prepends a deterministic recap of prior turns to each new request. In S2, after each turn, we run regression tests and trigger the gate only when $RPR_t < RPR_{t-1}$ and a last passing code suggestion exists. We then reject the current code suggestion, send the model a rollback prompt containing the failed-test count, the full last passing code suggestion, and the current-turn requirement, and allow exactly one retry. We retain the retry only when it passes at least as many benchmark tests as the rejected code suggestion; otherwise, the current code suggestion is kept. S3 applies the recap (S1) and the verification gate (S2) together. Each strategy is run on all 542 tasks with the full 8-turn chain on DeepSeek-V3 and Llama-3.1-8B. Table~\ref{tab:rq4-summary} reports $RPR_{T8}$, the fraction of tasks with regression, and average gate triggers per task. A gate trigger is counted whenever the current code suggestion lowers $RPR$ relative to the previous turn.

\begin{table}[t]
\caption{Mitigation results on 542 tasks. Gates denote the average number of verification gates triggered per task. $RPR_{T1}$ differences for DeepSeek reflect API non-determinism, not strategy effects, because mitigation starts at T2.}\label{tab:rq4-summary}
\centering
\small
\begin{tabular}{@{}llrrrl@{}}
\toprule
\textbf{Model} & \textbf{Strategy} & \textbf{$RPR_{T1}$} & \textbf{$RPR_{T8}$} & \textbf{\% With Regr.} & \textbf{Gates} \\
\midrule
\multirow{4}{*}{DeepSeek}
& Baseline   & 87.6 & 75.8 & 43.2\% & --- \\
& S1 (Recap) & 91.2 & 80.1 & 36.7\% & --- \\
& S2 (Gate)  & 91.1 & 87.9 & 12.5\% & 1.08 \\
& S3 (Both)  & 90.5 & 87.0 & 13.8\% & 1.21 \\
\midrule
\multirow{4}{*}{Llama-8B}
& Baseline   & 76.1 & 31.6 & 72.9\% & --- \\
& S1 (Recap) & 76.1 & 35.8 & 68.3\% & --- \\
& S2 (Gate)  & 76.1 & 47.3 & 54.8\% & 2.29 \\
& S3 (Both)  & 76.1 & 42.3 & 59.4\% & 2.33 \\
\bottomrule
\end{tabular}
\end{table}

\noindent\textbf{Finding 9. Verification Gate is the only strategy that produces large and stable mitigation gains across both models.} Under S2, DeepSeek-V3 improves from $RPR_{T8}=75.8\%$ to 87.9\% and reduces tasks with regression from 43.2\% to 12.5\%. Llama-3.1-8B improves from 31.6\% to 47.3\% and reduces tasks with regression from 72.9\% to 54.8\%. In contrast, Snowball Recap alone yields smaller improvements (DeepSeek-V3 80.1\%, Llama-3.1-8B 35.8\%), and combining recap with gating does not surpass S2 (S3 reaches 87.0\% and 42.3\%). This result indicates that online behavioral verification and rollback, rather than prompt restatement alone, is the main driver of mitigation in multi-turn coding.

\noindent\textbf{Finding 10. Mitigation effectiveness depends on model scale and correction frequency.} Under S2, the average number of triggered gates is 1.08 per task on DeepSeek-V3 and 2.29 on Llama-3.1-8B. The smaller model requires more than twice as many corrections and still produces lower-quality results. This suggests that Verification Gate removes a substantial share of regressions, but recovery after a failed turn remains harder for lower-capacity models.

\begin{tcolorbox}[colback=black!2,colframe=black!20,boxrule=0.4pt,arc=3pt,left=4pt,right=4pt,top=4pt,bottom=4pt]
\small
Our results show that Verification Gate is the only strategy that consistently and substantially mitigates multi-turn regression accumulation across both models.
Snowball Recap yields smaller gains, and combining the recap with the gate does not outperform the gate alone.
\end{tcolorbox}

\section{Discussion}\label{sec:discussion}

\subsection{Interpretation of the results}
%\noindent 
LLM-assisted programming is increasingly used in iterative settings where requirements evolve over multiple turns. In this setting, reliability is not only about satisfying the current request but also requires preserving requirements established in earlier turns. Our results show that this preservation requirement is where current models fail systematically. Across all six models, regression accumulates as conversations proceed. For five of the six models, the largest early losses appear at Turn~2 or Turn~3, where new validation or input-generalization logic must be integrated without breaking earlier-turn requirements. Stronger models degrade more gradually and finish at higher $RPR$ levels, whereas weaker models lose quality more sharply as requirements accumulate across later turns. The dominant failures are cross-turn incompatibilities rather than isolated single-turn mistakes.

\noindent We combine fixed multi-turn evolution chains, regression-oracle tracking at every turn, and a manually validated taxonomy of multi-turn bugs. This design separates first-turn code-suggestion errors from failures introduced by later requirement changes. It also makes model-level and turn-level comparisons reproducible.

\noindent The mitigation results extend this contribution from diagnosis to intervention. Snowball Recap helps when the model loses track of requirements established in earlier turns by restating them before each new turn. Verification Gate addresses a different problem: a later code suggestion may still break earlier-turn requirements even when those requirements are present in the conversation context. In that case, prompt restatement alone is not enough. Verification Gate is therefore the only strategy that yields consistent gains in both stronger and weaker models. Snowball Recap yields smaller gains, and adding it to Verification Gate does not improve on Verification Gate alone. This pattern suggests that the main problem in our setting is not simply loss of earlier-turn context, but failure to preserve earlier-turn requirements.

\noindent These results also suggest several directions for future LLMs for code generation. This points to a need for models that treat later turns less as isolated generation steps and more as constrained code updates over an evolving code state. More specifically, future systems may need stronger mechanisms for preserving earlier-turn requirements, checking compatibility between new and existing logic, and deciding when to revise from a last known-good code suggestion instead of continuing from a faulty one. Our mitigation results further suggest that these capabilities should not be left entirely to prompting. They are likely to benefit from tighter integration between code generation, regression testing, and turn-level acceptance checks.

\subsection{Implications}

\noindent (1) \textbf{Implication for evaluation design.} Single-turn scores are insufficient to characterize reliability in conversational programming. Evaluation protocols should track preservation across turns and include regression-aware metrics, because turn-local correctness can coexist with failure to preserve requirements established in earlier turns. In practice, this means that benchmark design should pair evolving requirements with stable regression oracles, so that each new turn is evaluated against both new functionality and earlier-turn requirements. Without this structure, evaluations can overestimate model robustness in maintenance-like workflows.

\noindent (2) \textbf{Implication for failure analysis.} Cross-Turn Conflict is the major failure class, with Over-Guarding and Semantic Collision accounting for most manually labeled failures. This suggests that future analyses should focus on how new constraints and new logic interact with requirements established in earlier turns, rather than attributing multi-turn failure primarily to context loss. A direct consequence is that failure taxonomies and debugging workflows for multi-turn coding should represent interaction faults explicitly, for example, incompatible validation changes, conflicting assumptions between turns, or integration breakage after local edits.

\noindent (3) \textbf{Implication for tool construction.} Tooling should prioritize verification-aware interaction loops, including turn-level regression checks and rollback-based retry from a known-good state. Our results indicate that this pattern is more effective than recap-first prompting, especially when requirements continue to evolve over turns. For deployment, this implies that practical assistants should treat each new code suggestion as a candidate change that requires acceptance checks, rather than assuming that additional prompt context alone will preserve earlier-turn requirements.

\subsection{Threats to Validity}
% In this section, we discuss the threats to internal, construct, and external validity of our study.
\noindent \textbf{Internal validity.} Our benchmark uses a fixed 8-turn sequence and controlled prompt templates. This improves comparability but cannot represent all natural conversation patterns. We mitigate this by applying the same protocol to all models, adding stratified analyses, and reporting initially solved-task results. We also set the temperature to 0.0 for reproducibility, while acknowledging that stochastic decoding in deployment may behave differently.

\noindent \textbf{Construct validity.} We measure preservation with turn-level regression pass rate ($RPR$) against benchmark tests. This captures tested behavioral preservation but not all quality dimensions, such as readability or untested edge cases. To reduce measurement artifacts, we reuse benchmark-validated tests and apply deterministic adaptation for interface restructuring turns. For taxonomy labeling, we use a frozen codebook, independent four-annotator coding, adjudication, and substantial agreement ($\kappa=0.799$).

\noindent \textbf{External validity.} Our tasks are Python function-level problems from HumanEval+ and MBPP+, so effect sizes may not directly transfer to other languages, repository-scale maintenance, or human-in-the-loop settings. We partially broaden coverage by evaluating six models from both the commercial and open-source families. Further replications on larger repositories and additional languages are needed.

\section{Conclusion}\label{sec:conclusion}

LLM-assisted programming is increasingly used in iterative software development workflows where requirements evolve over multiple turns. In this setting, reliability depends on whether newly generated code preserves behavior that was correct in earlier turns. This paper investigates the preservation problem through regression accumulation in multi-turn coding conversations.

We conduct a controlled empirical study on 542 tasks, six models, and 26,016 turn instances. We provide a reproducible benchmark protocol for turn-level regression evaluation, a taxonomy of multi-turn regression bugs from a four-annotator analysis of 384 stratified failures. We also evaluate interaction-level mitigation strategies based on recap and verification gate mechanisms.

Our findings show that regression accumulation appears across all evaluated models. We find that Cross-Turn Conflict is the major failure class, and that Verification Gate is the only mitigation strategy that yields consistent gains across both stronger and weaker models. Taken together, these results suggest that conversational code generation should be treated as an evolving maintenance process rather than a sequence of isolated prompt outcomes.


\section*{Data Availability}
Our benchmark (542 tasks $\times$ 8 turns), all experimental results across 6 models, bug annotations, and analysis scripts are available at \url{https://anonymous.4open.science/r/multiturn-code-bugs}.

\bibliographystyle{ACM-Reference-Format}
\bibliography{references}

\end{document}